\documentstyle[12pt]{article}
\input epsf
\begin{document}

\begin{center}
{\bf GLASSY DYNAMICS 
IN A MODEL WITHOUT DISORDER: 
SPIN ANALOG OF A STRUCTURAL GLASS}
\end{center}

LEI GU, BULBUL CHAKRABORTY\\
Department of Physics, Brandeis University, Waltham, 
MA 02254, 
USA,\\ bulbul@snow.cc.brandeis.edu
\begin{center}
{Paper presented at MRS Fall Meeting, Dec 1996}
\end{center}

ABSTRACT

We have analyzed  a non-randomly frustrated spin model 
which exhibits behavior remarkably similar to the 
phenomenology of structural glasses.  
The high-temperature disordered phase undergoes a strong 
first-order transition to a long-range ordered structure. 
Using Monte Carlo simulations, we have studied the behavior 
of the supercooled state by quenching
to temperatures below this transition temperature.  For a 
range of supercooling, the system remains ergodic and 
exhibits dynamics characteristic of supercooled liquids.
Below a certain characteristic temperature, however,  the 
system freezes
into a ``glassy" phase.    In this phase, the system is 
non-ergodic and
evolves through a distribution of traps characterized by a 
power-law
distribution
of trapping times.  This change in the dynamic 
behavior is concurrent with the appearance of a shear 
instability.

INTRODUCTION

Understanding of glassy dynamics and the glass transition 
remains as one of
the significant challenges in condensed matter physics.  
Recent experiments on
glasses [1,2] have led to some interesting 
observations regarding
the glass transition in structural glasses and indicated 
some similarities between
spin glasses and structural glasses [2].  
On the
 theoretical side, a
lot of interest has been focused on spin models without 
quenched-in disorder
which exhibit glassy dynamics [3].  The glassy 
dynamics in these
models, in some instances, have been related to the 
behavior of equivalent
spin-glass like models [3].  Frustration is a 
key concept in theories
of structural glasses such as the curved-space pictures 
of metallic
glass [4].  The two concepts taken together, 
glassy dynamics in
non-random spin systems 
and frustration as a key to 
glassy behavior, suggest
that frustrated spin systems could play a role in our 
quest for an understanding
the nature of structural glasses [5].  The 
model discussed in this paper belongs to this category.

It has been shown [6,7] that the 
frustration of a triangular-lattice Ising
antiferromagnet can be removed by elastic distortions.  
In the deformable lattice, the
antiferromagnetic couplings between nearest-neighbor 
spins depend on their
separation.   In
the simplest model, considered here, only uniform 
distortions of the lattice are allowed and
these are characterized by the relative changes 
$e_{1}~, e_{2}$ and $ e_{3}$ of the three
nearest-neighbor bonds lengths.  In equilibrium, 
there is a strong first-order transition
from the paramagnetic phase to a ``striped'' structure.  
The striped phase has alternating rows of up
and down spins and a frozen-in shear distortion 
with one set of bonds (between parallel
spins) expanded
and the other two (between antiparallel 
spins) contracted.   Monte Carlo
simulations have been used to demonstrate that this 
picture remains virtually unchanged when
fluctuations in bond-lengths are allowed [7].

A strong first-order transition accompanied by a shear 
distortion is reminiscent of the
density functional theory of freezing, where volume 
change plays a similar role, and it seems natural to 
ask how the supercooled spin system behaves.
Upon quenching the system from 
the disordered phase 
to the ordered regime,
a phenomenology remarkably similar to structural
glasses was observed.
Monte Carlo simulations were used to study the dynamics 
following instantaneous  quenches from a high-temperature 
disordered phase to a range of
temperatures below the ordering transition (which was 
strongly first-order). The dynamics employed was standard 
spin-exchange dynamics extended to include
moves which attempt global changes of the shape and size 
of the box [7].
These global changes were 
attempted after a complete sweep of all the spins in the 
lattice.  Three system sizes, 32x32, 48x48 and 64x64
were used to investigate the finite size dependence of
physical quantities.

DISCUSSION AND RESULTS

These simulations have 
lead to an intriguing picture of the 
supercooled state which could shed
some light on the nature of the structural glass 
transition.  Above a certain characteristic temperature, 
$T^*$, the relaxation of the supercooled state can be 
described reasonably well by a stretched
exponential behavior although there are some deviations.  
The time scale for
nucleation of the striped phase
is extremely long and the 
system behaves as if
it is in equilibrium in the metastable disordered phase.   
There is a volume
change of the lattice but the shear distortion fluctuates 
around zero, as seen
in Fig 1.  Fig. 1 also shows the time evolution of the 
energy per spin.  For $T
>T^*$,  the plateaus in these plots indicate the existence 
of shallow traps which are sampled by the system as it
explores the configurations available to it within the 
large trap or valley that
is the disordered
phase with zero shear distortion.
Below $T^*$, the
behavior of the system becomes non-ergodic.  
Time-translational invariance is
lost and there is continuous evolution or ``aging'' of 
the system.  The plots in Fig 1 show aging of the shear 
distortion and the energy per spin.
The system is no longer trapped within the large valley 
characterizing the disordered phase.   There is a clear 
evolution towards lower energy states but
this evolution is interrupted by trapping into states 
with 
varying well depths.
The evidence from simulation points towards a power-law 
distribution of the
trapping times in these wells.  If the power law is weak 
enough, this distribution could lead to aging since the 
system can always find a trap with a
trapping time comparable to the observation time 
[8].   

\begin{figure}[t]
\epsfxsize=5.4in \epsfysize=3.5in
\epsfbox{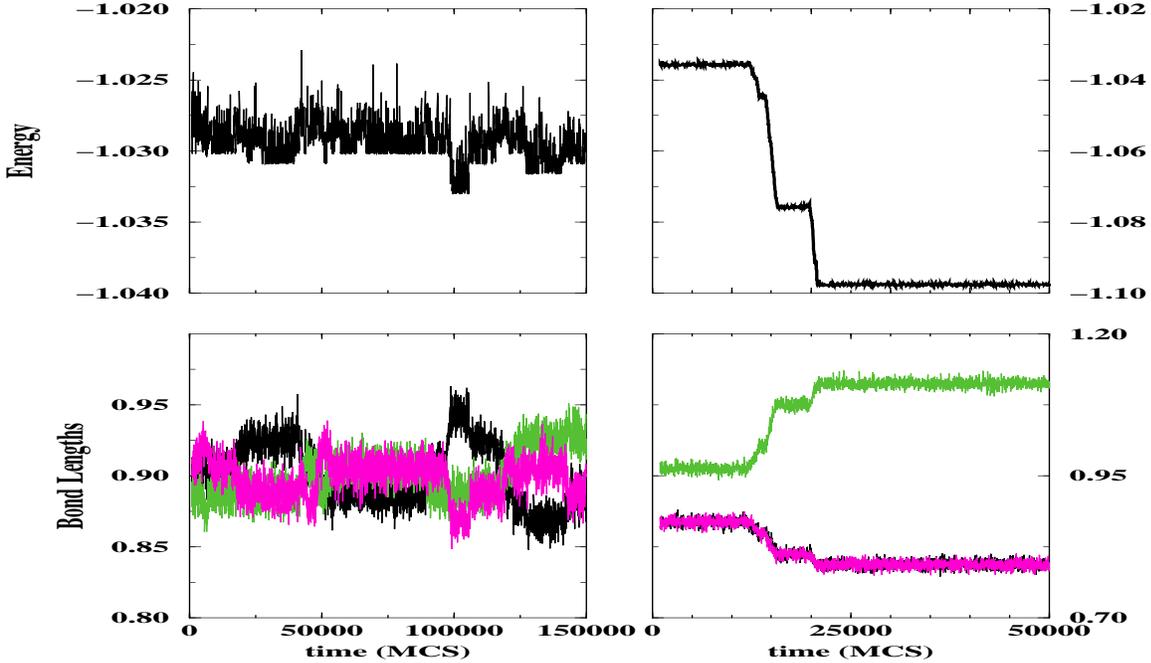}
\caption{Typical Monte Carlo trajectories showing the 
energy per spin  and the three bond 
lengths ($1+ e_{\alpha}$)
 vs. time. 
The left-hand panel shows data for a temperature 
$T>T^*$ and the right
hand panel shows data for $T<T^*$.  A shear distortion is indicated
by $e_1 >0$ and $e_2~, e_3 <0$ in the right-hand panel for $T<T^*$.  In the 
left-hand panel, the bond lengths are fluctuating about a common value, 
and there is no 
shear distortion.
The energy of the completely ordered striped phase 
is $-1.27$.}  
\label{fig1}
\end{figure}

The onset of
aging in this model is concurrent with the disappearance 
of the well
characterized by zero shear, or equivalently, with the 
appearance of a shear
instability.   It is known that this model has a shear 
instability at low enough
temperatures [6].  
This instability is normally 
not seen because the
first-order transition intervenes at a higher temperature. 
What is a true
instability in this mean-field model could become a narrowly 
avoided critical
point [5] when fluctuations of the bond lengths are included.  
This would, however,
not lead to any drastic changes in the scenario.  In 
the mean-field model
considered here, the supercooled state looses its stability 
at $T^*$.  This, by
itself, would have been uninteresting since it would have 
simply implied that
the nucleated path to the striped phase would be replaced 
by a continuous
ordering path.  What makes the dynamics non-trivial in 
this model is the existence
of intermediate states which cannot be avoided and are 
traversed only by 
barrier-hopping.  Therefore, the instability of the 
supercooled phase leads to
the appearance of a glassy phase rather than the 
striped phase.   
 
One of the characteristics of glass is the lack of 
ergodicity.  A measure of
broken ergodicity
can be obtained from examining the 
fluctuation metric [9].  This quantity, 
${\Omega}(t)$, measures the integral of
the auto-correlation function of the fluctuating 
quantity and is expected to
decay as $1/t$ for ergodic systems where the autocorrelation 
function decays to
zero.  Fig 2 shows the inverse of the fluctuation metric 
for four different
quench histories at temperatures above and below $T^*$.  
Above $T^*$ the system
is clearly ergodic and there is no dependence on quench
history.  This is far
from true for temperatures below $T^*$.   The dependence 
on quench history shows
that the evolution of the system takes place in a 
rugged landscape.  This
statement could be made more accurate by analyzing the 
probability distribution of energy, $P(E)$ in the two 
different temperature regimes. 

\begin{figure}[htbp]
\epsfxsize=5.4in \epsfysize=2.75in
\epsfbox{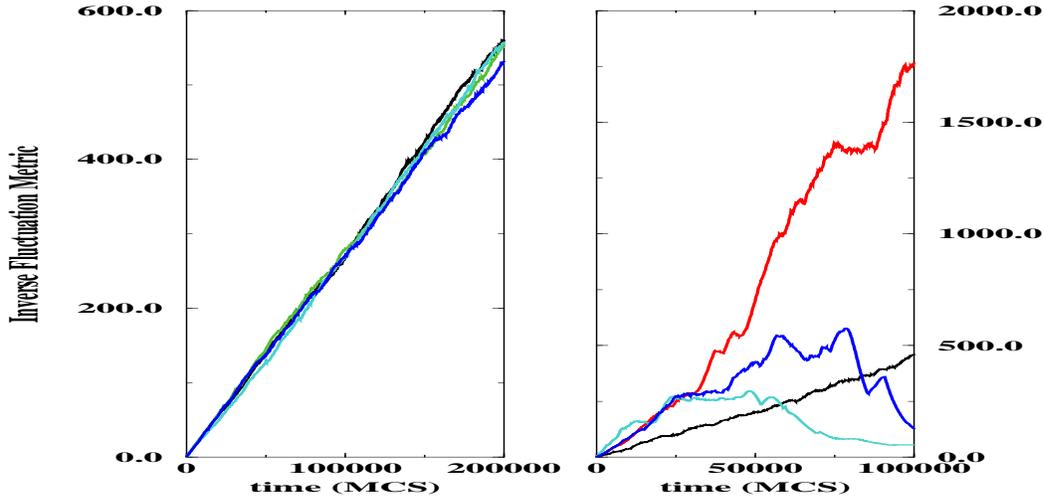}
\caption{Plot of the inverse of the fluctuation metric 
for four different Monte
Carlo runs.  The
left panel shows data for $T>T^*$ and the 
right panel shows
data for $T<T^*$.}  
\label{fig2}
\end{figure}

In trying to deduce a picture of the free-energy surface 
from the simulation
data, it is useful to characterize a valley by a region 
of phase space in which
the system spends a macroscopic amount of time, much 
larger than the microscopic
time scale for spin exchange.  These valleys can then be 
characterized by local
averages such as the energy per spin or the shear 
distortion.  
In
 equilibrium, the probability of finding a system 
in a given valley 
characterized, for example, by energy $E_{\alpha}$ 
is proportional to $\exp
(-{\beta}F_{\alpha})$ where $F_{\alpha}$ is the free 
energy characterizing that
state.  Since the system is observed to reach
quasiequilibrium in the valleys,
one can 
obtain a picture of the
free energy landscape by measuring the energy histogram $P(E)$ and
examining the behavior of $-\ln P(E)$.  
These free energy surfaces for
$T>T^*$ and $T<T^*$ are shown in Fig 3.  The difference 
between the two
justifies the picture presented earlier
.  At the higher 
temperatures, there is
a large valley with a subvalley structure.  The system, 
for large enough system
sizes, never escapes this valley and for all intents and 
purposes, is in
equilibrium.  The breaking of ergodicity here is akin to 
that in a ferromagnet
and does not lead to glassy dynamics.  Simulations at 
higher temperatures show
that the internal structure of this valley is not 
present at high enough
temperatures.  This indicates that as the temperature is
lowered, 
the subvalleys become deeper in relation to 
the overall depth of the
large valley.  The relaxation time within this large 
valley should be affected
by the changing internal structure and, it is possible  
that these
structures lead to the unusual dynamics observed in 
supercooled liquids above
the glass transition [2].  This aspect is 
being investigated at present.
The large valley is absent from the free energy surface for $T<T^*$.
The trend
of the subvalleys becoming
 deeper in relation to the 
overall depth of the large
valley has led to the disappearance of this overall 
structure.  The system is
now free to explore a larger region of phase space.  
However, the subvalleys,
which have now acquired the role of the primary valleys, 
trap the system over all different time scales.  

\begin{figure}[thbp]
\epsfxsize=5.4in \epsfysize=2.75in
\epsfbox{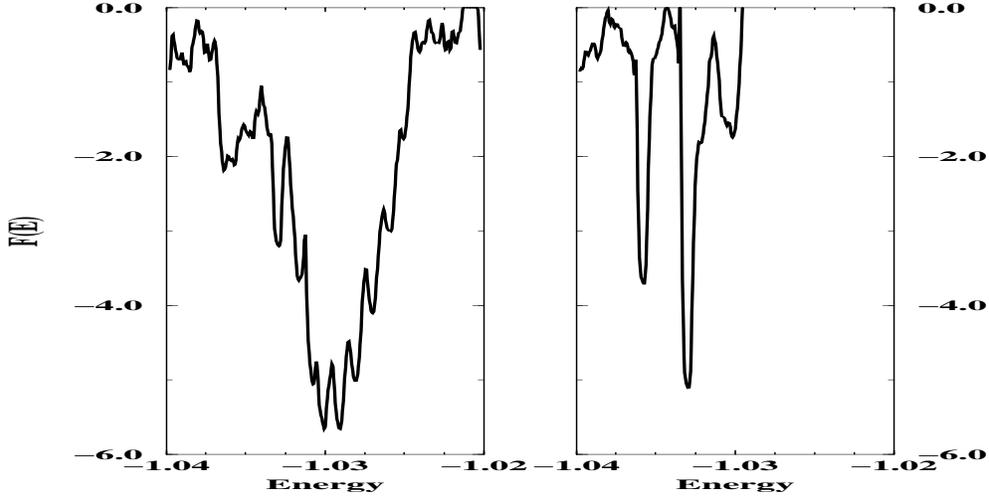}
\caption{The topology of the free energy surface obtained 
from $-{\ln} P(E)$.
This prescription gives
us the free energy differences 
between different states
at a given temperature.
Hence, it can provide a description of the topology but 
not absolute magnitudes
or comparison between different temperatures. The left 
panel is for $T>T^*$ and
the right panel for $T<T^*$.}
\label{fig3}
\end{figure}

A theory of aging in glasses has been proposed which is  
based on a broad
distribution of trapping times [8].  
This distribution can be obtained from 
the energy versus
time
data obtained in the simulations.  
A histogram of trapping
times at $T < T^*$ (Fig. 4) shows a power law 
distribution and is
consistent  with
the class of power laws discussed in the theory of 
aging [8].
These  power law distributions imply that the {\it average} 
trapping time
diverges and the system cannot explore all of phase space 
in a finite amount of
time. This phenomenon has been referred to as weak 
ergodicity
breaking [8].

\begin{figure}[bhp
t]
\epsfxsize=5.4in \epsfysize=2.75in
\epsfbox{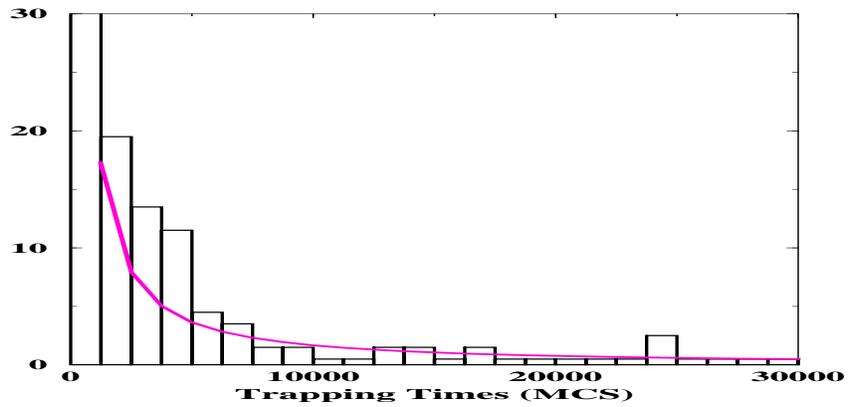}
\caption{Histogram of trapping times, $P({\tau})$ 
for $T<T^*$. The
solid line is a fit to a power law, $P({\tau}) = 
({\tau})^{(-1.2)}$.
This distribution
law implies a divergent average, $<{\tau}>$.}
\label{fig4}
\end{figure}

There seems to be a correlation between the trapping 
times in the various wells
and the shear distortion characterizing these state.  
The larger the shear
distortion the longer is the trapping time.  An 
examination of the morphologies
of the states in these wells shows that there are 
regions of locally ordered
striped phases in these wells.  These locally ordered 
regions are separated by
disordered regions.  The disordered regions get 
squeezed into smaller and
smaller channel as the system evolves towards lower-energy valleys.
The trapping time
increases as these channels get smaller.  The aging 
process, therefore, leads to
valleys with larger shear strains and smaller 
disordered regions.  However
,
since this leads to the system getting trapped in these 
wells for longer and
longer times, the aging process, in effect, never stops.

The height of the barrier, or equivalently, well depth of the large disordered valley
was found to scale with the systems size.  For $T<T^*$, it is more meaningful to talk
about the distribution of well depths or trapping times.  Preliminary results
indicate that these distribution extends to longer times as the system size 
increases.  What is clear, in this temperature regime, is that some barriers do not
scale with system size.

CONCLUSION

The stepwise relaxation seen in Fig 1 is akin to the 
observed
relaxation in the multispin models [10].  
As far as the equilibrium properties are concerned, 
the deformable lattice model
considered in this paper, can be mapped onto a pure spin 
model with pair
interactions which are short-ranged and four-spin 
interactions which are long
ranged.
  It is therefore not surprising that there 
are similarities between
these models.  The basic driving force behind the 
complexity observed in the
deformable lattice model is the frustration inherent 
in the triangular lattice.
Coupling to the lattice removes the degeneracies and 
in its place introduces
structure on small and large energy scales.   The 
instability to shear owes its
existence to the degeneracy of the triangular lattice 
groundstate [6].
The coupling between local variables
which are 
frustrated and global variables
which are capable of removing the frustration seem to 
be the essential
ingredients of this model.  It is possible that this 
feature is also present in
real glasses and is responsible for the remarkable 
similarity of the
phenomenology of our toy model with real systems.

ACKNOWLEDGEMENT

The authors would like to acknowledge useful discussions with N. Gross, Bill Klein
and Royce Zia. This work has been supported in part by DE-FG02-ER45495.

REFERENCES

1. C. A. Angell, Proc. Nat. Acad. Sci. {\bf 92}, 6675 (1995).\\
2. Narayanan Menon and Sidney R. Nagel, Phys. Rev. Lett. {\bf 74}, 1230 (1995) ;
D. Bitko {\it et al}, 
Europhys. Lett {\bf 33}, 489 (1996).\\
3. J. P. Bouchaud and M. Mezard, J. Phys. I {\bf 4}, 1109 (1994).\\
4. For a recent review and simulation studies, see
T. Tomida and T. Egami, Phy. Rev. B {\bf 52}, 3290 (1995).\\
5. S. A. Kivelson {\it et al}, J. Chem. Phys {\bf 101}, 2391 (1994).\\
6. Z. Y. Chen and M. Kardar, J. Phys. C {\bf 19}, 6825 (1986).\\
7. Lei Gu, Bulbul Chakraborty, P. L. Garrido, Mohan Phani and
J. L. Lebowitz, Phys. Rev. B {\bf 53}, 11985 (1996).\\
8. J. P. Bouchaud, J. Phys. I {\bf 2}, 1705 (1992).\\
9. D. Thirumalai and Raymond D. Mountain, Phys. Rev. B{\bf 47}, 479 (1993).\\
10. W. Krauth and M. Mezard,
Z. Phys. B {\bf 97}, 127 (1995).\\
\end{document}